\documentclass{Interspeech2024}
\usepackage{algorithm}
\usepackage{algpseudocode}
\usepackage{subcaption}

\interspeechcameraready

\title{Frame-Wise Breath Detection with Self-Training: An Exploration of Enhancing Breath Naturalness in Text-to-Speech}

\name[affiliation={1,2}]{Dong}{Yang}
\name[affiliation={1}]{Tomoki}{Koriyama}
\name[affiliation={2}]{Yuki}{Saito}

\address{
  $^1$CyberAgent, Inc., Japan, $^2$The University of Tokyo, Japan}
\email{
ydqmkkx@gmail.com
}


\begin{document}
\setlength{\abovedisplayskip}{5pt} 
\setlength{\belowdisplayskip}{5pt} 
\setlength\floatsep{8pt} 
\setlength\intextsep{6pt} 
\setlength\textfloatsep{6pt} 
\setlength{\dbltextfloatsep}{6pt} 
\setlength{\dblfloatsep}{5pt}

\maketitle


 
\begin{abstract}
Developing Text-to-Speech (TTS) systems that can synthesize natural breath is essential for human-like voice agents but requires extensive manual annotation of breath positions in training data. To this end, we propose a self-training method for training a breath detection model that can automatically detect breath positions in speech. Our method trains the model using a large speech corpus and involves: 1) annotation of limited breath sounds utilizing a rule-based approach, and 2) iterative augmentation of these annotations through pseudo-labeling based on the model's predictions. Our detection model employs Conformer blocks with down-/up-sampling layers, enabling accurate frame-wise breath detection. We investigate its effectiveness in multi-speaker TTS using text transcripts with detected breath marks. The results indicate that using our proposed model for breath detection and breath mark insertion synthesizes breath-contained speech more naturally than a baseline model.

\end{abstract}

\vspace{-5pt}
\section{Introduction}
\vspace{-3pt}
With the ongoing advancement of TTS systems, the pursuit of naturalness in synthetic speech has made nuanced aspects like Pause-Internal Phonetic Particles (PINTS)~\cite{pints} gain more attention. PINTs are the elements within speech pauses, such as silence, breath~\cite{previous0, baseline1, baseline2}, tongue clicks, filled pauses~\cite{filled0, filled1}, and laughter~\cite{laughter0, laughter1}. Among these elements, breath is important for perceived naturalness in synthetic speech~\cite{importance, previous0, baseline1} but is often less naturally represented, a limitation that has been exploited in deepfake detection~\cite{deepfake}. 

To enhance the naturalness of breath in synthetic speech, a widely adopted strategy is to detect breath sounds in speech data and insert corresponding breath marks into the text data for TTS training~\cite{previous0, baseline2}. Then, the breath marks are predicted and inserted during the inference phase~\cite{baseline1}. However, conventional breath detection methods come with several limitations. They generally fall into two categories: one relies on acoustic feature analysis for rule-based detection~\cite{zcr, previous0}, and the other manually annotates the breath data and trains machine learning models for detection~\cite{zcr1, baseline}. The former, though not requiring manual data annotation, tends to compromise on accuracy; while the latter, despite achieving higher accuracy, fails to yield a universal and robust detection model without extensive annotation across large speech corpora. To our knowledge, these methods have only been applied in small speech corpora.

In this study, we primarily concentrate on breath detection to overcome existing limitations. We introduce an innovative training approach that eliminates the need for manual annotation of breath in the training dataset. Our approach is a hybrid one that utilizes the advantages of previous rule-based and machine-learning-based approaches with a self-training method. Moreover, we propose a powerful frame-wise breath detection model trained on a large multi-speaker speech corpus. Our method starts with a comprehensive acoustic feature analysis for breath extraction. We not only verify the effectiveness of several features from previous works but also put forward additional features. Based on these features, we develop a rule-based approach to automatically extract and annotate a sufficient number of breath sounds within the training set, aiming to train breath detection models. Our proposed
breath detection model incorporates Conformer~\cite{conformer} and bidirectional long short-term memory (BiLSTM). It is optimized with downsampling and upsampling modules, enabling frame-wise detection while reducing computational costs. Through a self-training methodology~\cite{selftraining0, selftraining1}, the model is trained efficiently with both annotated and unannotated data, leading to superior performance. Experimental results demonstrate that our proposed model significantly outperforms the baseline model in breath detection. When training the TTS model with speech and text transcripts augmented with detected breath marks from the proposed model, the synthetic breath sounds exhibit enhanced naturalness. Speech samples and code are available\footnote{\url{https://ydqmkkx.github.io/breath-detection/}}.

The main contributions of this paper are:
\begin{itemize}

    \item We propose two novel acoustic features related to the Mel-spectrogram, demonstrating effectiveness in breath detection.
    \item Our proposed model achieves frame-wise breath detection with finer time resolution compared to the baseline model's frame-level detection.
    \item Combining a rule-based approach with a self-training method, we efficiently train the breath detection model without the need for manual annotation of training data.
    \item Through multi-speaker TTS training, we observe that our proposed method enables the synthesis of breath sounds for speakers whose training data lacks breath sounds.
    \item This research lays the groundwork for breath modeling in TTS and provides a foundation for breath position detection.

\end{itemize}

\vspace{-5pt}
\section{Methods}
\vspace{-3pt}
\subsection{Data, acoustic features, and annotation}
\label{section: annotation}
\vspace{-3pt}
In this study, various subsets of the LibriTTS-R~\cite{librittsr} corpus are allocated for both breath detection and TTS experiments. It is a multi-speaker speech corpus derived from audiobooks and processed with speech restoration. Notably, although breath sounds consist of inflation and exhalation, exhalation is less in human speech and rare in LibriTTS-R corpus. Our study does not distinguish them specifically. 

Montreal Forced Aligner (MFA)~\cite{mfa} is utilized for aligning text transcripts with speech and detecting pauses. These pauses include several kinds of PINTs, mainly silence, breath, and tongue clicks. To evaluate our breath detection methods, we manually annotated the breath sounds within these pauses. The annotation was conducted on randomly selected sentences from the ``dev-clean'' and ``test-clean'' subsets of LibriTTS-R, thereby forming respective validation and test sets of breath detection experiments, the statistics of which are detailed in Table~\ref{tab: valid/test stat}.

\begin{table}[t]
  \caption{Statistics of the validation and test sets in breath detection experiments (manually annotated).}
  \label{tab: valid/test stat}
  \centering
  \vspace{-3mm}
  \begin{tabular}{ l r r r }
    \toprule
     & \textbf{Sentences} & \textbf{Pauses} & \textbf{Annotated breath} \\
    \midrule
    Validation set & 520 & 2049 & 400 \\
    Test set & 455 & 2051 & 480 \\
    \bottomrule
  \end{tabular}
\end{table}

\begin{table*}[ht]
  \caption{Thresholds and performance of automatic annotation method.}
  \label{tab: thresholds}
  \centering
  \vspace{-3mm}
  \begin{tabular}{l | c c c c | r r}
    \toprule
   \textbf{Class} & \textbf{Duration} & \textbf{Max(VMS)} & \textbf{Max(ZCR)} & \textbf{NA-VMS} & \textbf{Precision} & \textbf{Recall}\\
    \midrule
    Breath & $>300$ ms & $>150$ & $>1\times10^{-4}$ & $>0.6$ & 0.982 & 0.450\\
    Non-breath & - & $<150$ & $<5\times10^{-5}$ & - & 1.000 & 0.111\\
    \bottomrule
  \end{tabular}
\end{table*}

\begin{figure}[t]
  \centering
  \includegraphics[width=1.0\linewidth]{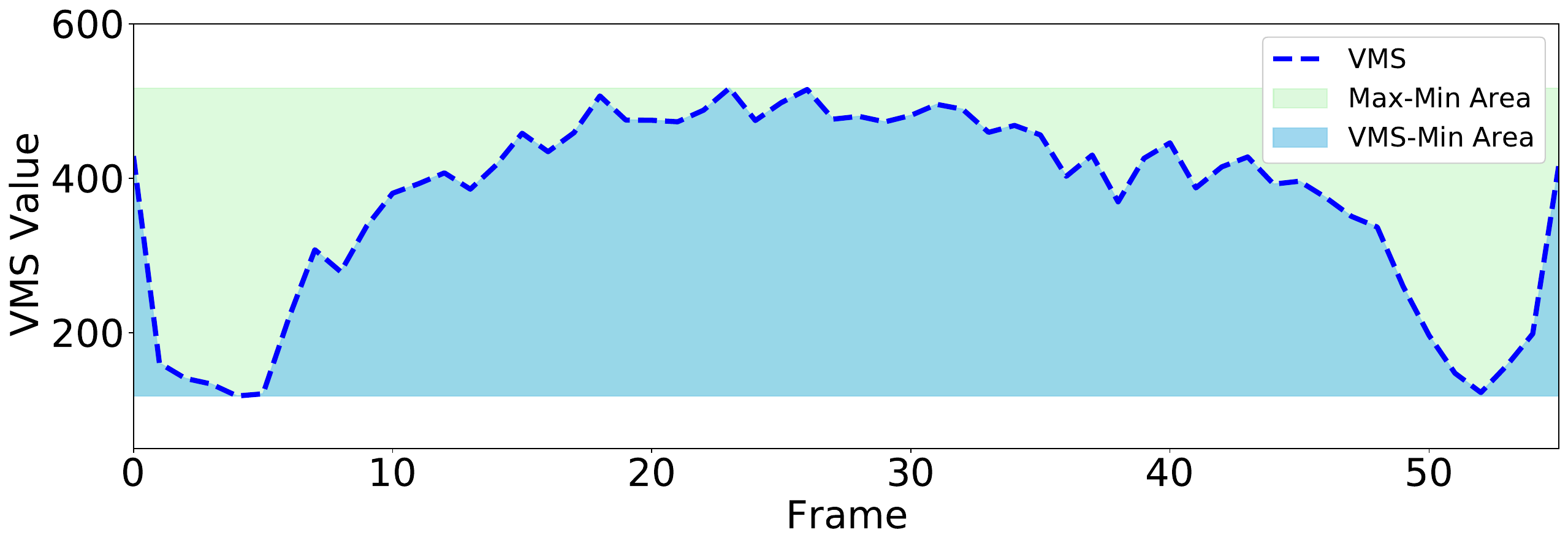}
  \vspace{-6mm}
  \caption{VMS curve within a pause segment.}
  \label{fig: vms}
\end{figure}

To train the breath detection model, we expect to extract and annotate an adequate amount of breath sounds automatically from the MFA-recognized pauses with high precision. Drawing from related works on acoustic feature analysis of breath, our preliminary experiments indicated the effectiveness of duration and Zero-Crossing Rate (ZCR)~\cite{zcr, zcr1, baseline, baseline1}. However, these two features alone are not sufficient for our purposes, leading us to put forward two additional acoustic features: Variance of Mel-Spectrogram (VMS) and Normalized Average of VMS (NA-VMS). The four features are detailed as follows:
\begin{enumerate}
\item \textbf{Duration}: Previous research indicates that breath sounds generally exhibit longer durations compared to other types of pauses~\cite{previous0, previous1}. Accordingly, we exclude pauses shorter than 300ms (brief pauses~\cite{phrasing}) for the extraction of breath data, which primarily consist of brief silence and noise.

\item \textbf{ZCR}: Defined as the number of times the audio signal changes its sign~\cite{zcr}. The computation of ZCR within a sliding window is depicted in Equation~\ref{equation: zcr}, where $N$ represents the window length and $X=\{x[n]\}_{n=0}^{N-1}$ denotes the sampled audio signal within a window.
\begin{equation}
  \text{ZCR}(X) = \frac{1}{N-1} \sum_{n=1}^{N-1} 0.5 | \text{sgn}(x[n]) - \text{sgn}(x[n-1]) |
  \label{equation: zcr}
\end{equation}

\item \textbf{VMS}: The variance of the Mel-spectrogram in its frequency domain. The maximum values of ZCR and VMS assist in distinguishing silence from other types of pauses because silence tends to have lower ZCR and VMS.

\item \textbf{NA-VMS}: The mean of min-max normalized VMS values over a range of frames, as defined in Equation~\ref{equation: navms} where $V=\{v[f]\}_{f=0}^{F-1}$ represents the VMS values across $F$ frames. It quantifies the ratio of the VMS-Min Area to the Max-Min Area as shown in Figure~\ref{fig: vms}. For pause segments, NA-VMS is effective for distinguishing breath sounds from tongue clicks because tongue clicks typically show a short peak in VMS and ZCR but a low NA-VMS.


\begin{equation}
  \text{NA-VMS}(V) = \frac{1}{F} \sum_{f=0}^{F-1} \frac {v[f] - \min(V)} {\max(V) - \min(V)}
  \label{equation: navms}
\end{equation}

\end{enumerate}

We used a sampling rate of 22,050~Hz for audio signals. Mel-spectrogram was generated and log-scaled by librosa~\cite{librosa}, configured with 256 Mel bands, 256 window length, and 128 hop length. ZCR was extracted with the same window length and hop length. Table~\ref{tab: thresholds} lists the thresholds of acoustic features for pause annotation, which were determined based on observations of the training data and then fine-tuned on the validation set. Using this rule-based breath detection, we achieved a precision of 0.982 in the test set. Furthermore, we extracted the non-breath data that is mainly silence, with an extraction precision of 1.00 in the test set. The non-breath data facilitates the training of the breath detection model that is explored in Section~\ref{breath detection experiments}. To utilize the remaining unclassified pauses, which constitute the majority of the pauses, we employed a self-training method in the training that is detailed in Section~\ref{self-training}.

\vspace{-3pt}
\subsection{Breath detection models}
\vspace{-3pt}
We explain two models that are evaluated in Section~\ref{experiment}.

{\bf Baseline model:} The breath detection model proposed by Sz{\'{e}}kely et al.~\cite{baseline} is the latest advancement in the field. This model has been utilized for spontaneous dialogue segmentation and spontaneous speech synthesis~\cite{baseline1, baseline2}. We implemented it as our baseline model, following its original description closely. The model consists of Convolutional Neural Network (CNN) layers, max pooling layers, and a BiLSTM layer. It takes Mel-spectrogram and ZCR as inputs, enabling frame-level detection. These features are extracted using a window length of 20~ms and a hop length of 2.5~ms. Due to the downsampling by the pooling layers, the model outputs a predictive probability of breath every 20 frames and achieves a detection resolution of 50~ms.

\begin{figure}[t]
  \centering
  \includegraphics[width=0.8\linewidth]{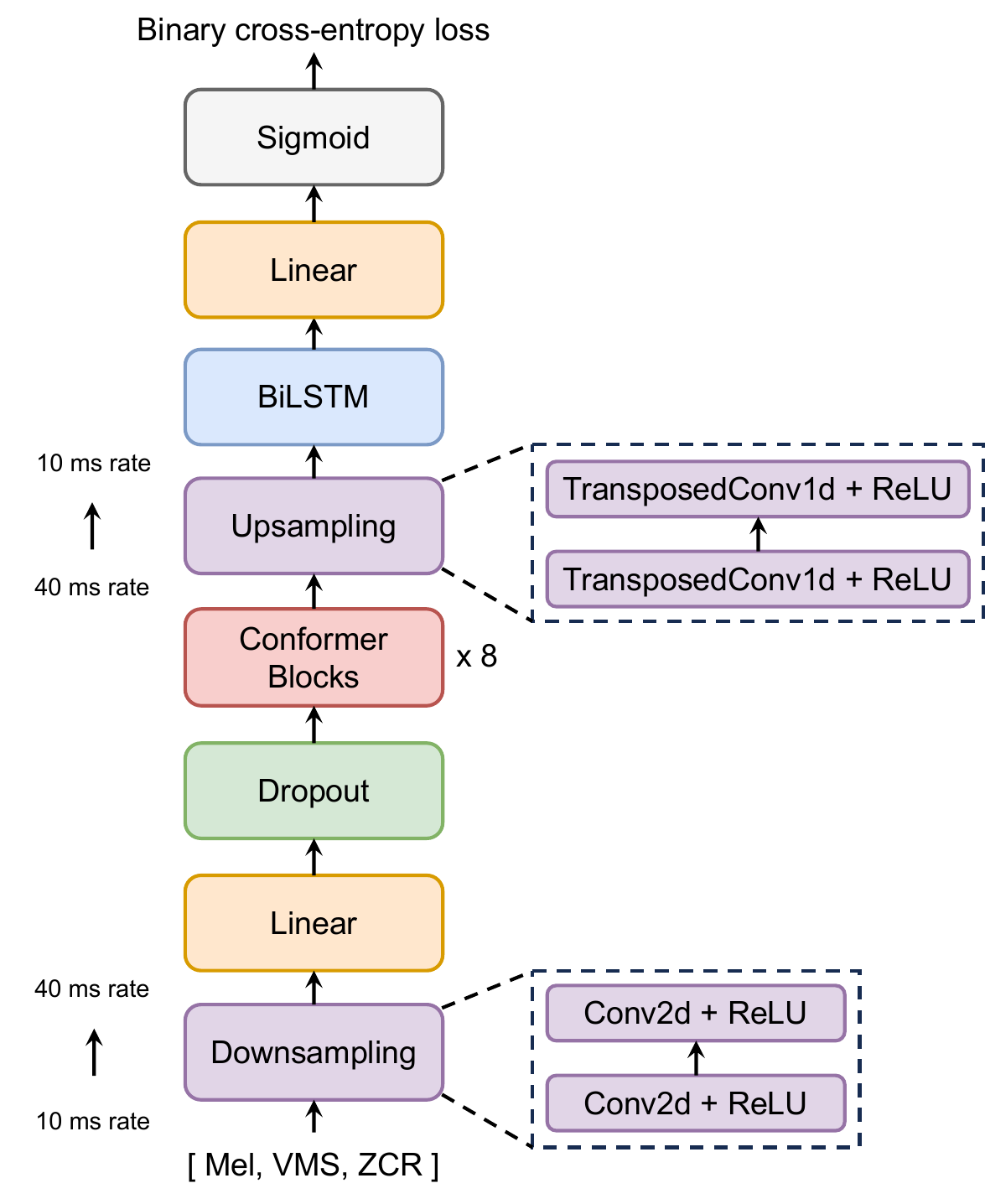}
  \vspace{-3mm}
  \caption{Architecture of proposed model.}
  \label{fig: model}
\end{figure}

{\bf Proposed model:} In light of the compelling performance of the Conformer architecture~\cite{conformer} in Automatic Speech Recognition (ASR), our proposed model incorporates Conformer blocks within the encoder and utilizes a BiLSTM layer as the decoder. While the downsampling module preceding the Conformer blocks is instrumental in local feature extraction and computational efficiency, it reduces the detection resolution as observed in the baseline model. To counteract this, we introduce an upsampling module after the Conformer blocks to reconstruct the time dimension and enable frame-wise detection. This method has been used for target alignment in ASR~\cite{deconv}. As shown in Figure~\ref{fig: model}, the downsampling module includes two 2D-CNN layers with a kernel size of 3$\times$3 and a stride width of 2$\times$2. Correspondingly, the upsampling module utilizes two transposed 1D-CNN layers with a kernel size of 3 and a stride width of 2 to recover the length of the hidden states. In terms of input, the Mel-spectrogram is integrated with not only ZCR but also VMS as two additional channels. The input features are extracted using a window length of 25~ms and a hop length of 10~ms, enabling the model to reach a detection resolution of 10~ms. In our experiments, 8 Conformer blocks were configured with 4 attention heads and a convolutional kernel size of 31. The hidden size of the model was 256 and the dropout rate was 0.1. The Sigmoid function was applied to output the predictive probability of breath for each frame and binary cross-entropy loss was used to train the model.

\vspace{-3pt}
\subsection{Self-training method}
\label{self-training}
\vspace{-3pt}

Let $X=\{x_{n,t}\}_{ 0 \leq n \leq N, t \in T_n}$ be the training set, where $N$ denotes the total number of speech utterances. Here, $x_{n,t}$ represents the $t$-th frame obtained from the $n$-th speech utterance through feature extraction. The label $y_{n,t}$ of $x_{n,t}$ is assigned according to Equation~\ref{equation: label} that involves four frame sets: all frames ($T_n \in T$), frames within MFA-recognized pauses ($P_n \in P$), frames within the breath set ($B_{n} \in B$), and frames within the non-breath set ($U_{n} \in U$). Notably, frames labeled with $-100$ do not contribute to the loss calculation:
\begin{equation}
 y_{n,t} =
\begin{cases}
    0 & {(t \in (T_n \setminus P_n) \cup U_n),} \\
    1 & {(t \in B_n),} \\
    -100 & {(t \in P_n \setminus (B_n \cup U_n)).}
\end{cases}
\label{equation: label}
\end{equation}

\begin{algorithm}
\caption{Self-training for breath detection models}
\begin{algorithmic}[1]
\State \textbf{Input:} $X$, $T$, $P \subseteq T$, $B \subseteq P$, $U \subseteq P$
\State $k \leftarrow 0$
\State $Y \leftarrow label(X, T, P, B, U)$ \Comment{Following Equation 3}
\State $D_{\theta}^{0}$ $\leftarrow$ initial detector trained on $(X, Y)$
\Repeat
    \State $k \leftarrow k+1$
    \State $\hat{B} \leftarrow D_{\theta}^{k-1}(X)>\alpha^k$  
    \State $\hat{U} \leftarrow D_{\theta}^{k-1}(X)<\beta^k$ 
    \State $Y \leftarrow label(X, T, P, B \cup \hat{B}, U \cup \hat{U})$ \Comment{Pseudo-labeling}
    \State $D_{\theta}^{k}$ $\leftarrow$ detector $D_{\theta}^{k-1}$ trained on $(X, Y)$
\Until{the performance of $D_{\theta}^{k}$ declines}
\State \textbf{Output:} $D_{\theta}^{k-1}$
\end{algorithmic}
\label{algorithm: self-training}
\end{algorithm}

The self-training process, as outlined in Algorithm~\ref{algorithm: self-training}, begins with the training of the initial detector $D_{\theta}^{0}$. In subsequent iterations, the breath and non-breath sets are augmented with pseudo-labels ($\hat{B}$ and $\hat{U}$). The pseudo-labels are generated based on the detector's predictions and dynamic thresholds ($\alpha^k$ and $\beta^k$). Specifically, the frames with a predictive probability above $\alpha^k$ are assigned to $\hat{B}$, while those with a predictive probability below $\beta^k$ are included in $\hat{U}$. These pseudo-labels are used to progressively refine the detector's training. The iterative training continues until the detector’s performance declines on the validation set, yielding the best-performing detector $D_{\theta}^{k-1}$.

\vspace{-5pt}
\section{Experiments and Results}\label{experiment}
 \vspace{-3pt}
\subsection{Breath detection experiments}
\vspace{-3pt}
We first conducted breath detection experiments to verify that our model achieves higher performance than the baseline.

{\bf Configurations:}\label{breath config}
The ``train-clean-100'' and ``train-other-500'' subsets of LibriTTS-R corpus were used as the training set. The breath and non-breath sets were extracted and annotated following the rule-based breath detection outlined in Section~\ref{section: annotation}. Speech signals were sampled at a rate of 16,000~Hz and processed to extract log-scaled Mel-spectrograms with 128 Mel bands, to serve as input features. The training was conducted on one NVIDIA A100 GPU with the AdamW optimizer~\cite{adamw}. The linear learning rate scheduler was utilized, with the initial 10\% training steps as the warm-up period. The learning rate increased from 0 to the peak during the warm-up period and then decreased to 0 linearly. During each training iteration, both baseline and proposed models were trained for 10 epochs with a batch size of 64 and a peak learning rate of $2 \times 10^{-5}$. 

The dynamic thresholds at the $k$-th iteration for pseudo-labeling, $\alpha^k$ and $\beta^k$, were adjusted through a classification task within the pauses of the validation set. Specifically, the detector $D_{\theta}^{k-1}$ output a probability for each pause frame, then the frame was classified as breath if its predictive probability was above $\alpha^k$ and as non-breath if below $\beta^k$. The target precision of classification for both breath and non-breath was initially set at 0.98 and decreased by 0.02 in each subsequent iteration. 

We also trained the baseline model with self-training for comparative analysis. In the initial training (i.e., training $D_{\theta}^{0}$ in Algorithm 1), we conducted ablation studies to explore the impact of incorporating ZCR and VMS as inputs, as well as using the non-breath set in training, on the performance of our proposed model. Furthermore, at iteration 1, we experimented with training the proposed model continuously without pseudo-labeling to investigate the effectiveness of self-training.

{\bf Results and analysis:}\label{breath detection experiments}
For this detection task, Intersection over Union (IoU) was adopted as the primary evaluation metric, complemented by precision and recall for more detailed analysis. The optimal predictive threshold was determined at the point where the model maximized IoU for breath detection in the validation set. This threshold was then applied to calculate these metrics in the test set. The experimental results are listed in Table~\ref{tab: results of breath detection}, which reveal the following findings:
\begin{itemize}
    \item From Table~\ref{tab: results of breath detection}(a), the proposed model consistently outperformed the baseline model, demonstrating its robustness. In addition, both of the two models achieved their peak IoU after the 3rd iteration of self-training, where the models were considered as the best-performing ones. Then the training finished after the 4th iteration, where both models improved the precision further but degraded IoU and recall a little.
    \item From Table~\ref{tab: results of breath detection}(b), the absence of either ZCR or VMS in the input, especially ZCR, significantly reduced the model's performance. Additionally, incorporating the non-breath set into the training was also proved to be critical. Furthermore, instead of improving, the proposed model's performance declined after the 1st iteration without pseudo-labeling, highlighting the significance of the self-training method.

\end{itemize}

\begin{table}[t]
  \caption{Results of breath detection experiments. Iter.: iteration of self-training.}
  \label{tab: results of breath detection}
  \centering
  \vspace{-2mm}

  \subcaption{Evaluation of self-training in baseline and proposed models.}
  \vspace{-1mm}
  \begin{tabular}{ l | r | r r r }
    \toprule
    \textbf{Model} & \textbf{Iter.} & \textbf{IoU} & \textbf{Precision} & \textbf{Recall} \\
    \midrule
    Baseline & 0 & 0.616 & 0.774 & 0.751 \\
     & 1 & 0.634 & 0.711 & \textbf{0.854} \\
     & 2 & 0.681 & 0.787 & 0.835 \\
     & 3 & \textbf{0.710} & 0.836 & 0.824 \\
     & 4 & 0.709 & \textbf{0.882} & 0.783 \\
    \midrule
    Proposed & 0 & 0.777 & 0.900 & 0.850 \\
     & 1 & 0.809 & 0.926 & 0.865 \\
     & 2 & 0.829 & 0.929 & 0.885 \\
     & 3 & \textbf{0.836} & 0.924 & \textbf{0.897} \\
     & 4 & 0.827 & \textbf{0.930} & 0.881 \\
    \bottomrule
  \end{tabular}

  \vspace{2mm}
  
  \subcaption{Ablation studies on proposed model.}
  \vspace{-1mm}
  \begin{tabular}{ l | r | r r r }
    \toprule
    \textbf{Model} & \textbf{IoU} & \textbf{Precision} & \textbf{Recall} \\
    \midrule
    Proposed (Iter. 0) & 0.777 & 0.900 & 0.850 \\
    ~~~~~~~~~w/o ZCR & 0.631 & 0.733 & 0.819 \\
    ~~~~~~~~~w/o VMS & 0.677 & 0.814 & 0.802 \\
    ~~~~~~~~~w/o non-breath & 0.702 & 0.785 & 0.869 \\
    \midrule
    Proposed (Iter. 1) & 0.809 & 0.926 & 0.865 \\
    ~~~~~~~~~w/o pseudo-label & 0.740 & 0.873 & 0.829 \\
    \bottomrule
  \end{tabular}
  
\end{table}

\vspace{-3pt}
\subsection{TTS experiments}\label{tts_experiment}
\vspace{-3pt}
We then evaluated the performance of our breath detection model in multi-speaker TTS experiments.

{\bf Configurations:}
 To effectively train the TTS models with sufficient pauses, the speech utterances in the ``train-clean-360'' subset of LibriTTS-R were segmented into clips ranging from 5 to 10 seconds for training and evaluation. We employed VITS~\cite{vits} as the backbone TTS model and utilized the baseline and proposed model (after the 3rd iteration in Table~\ref{tab: results of breath detection}) for breath detection. Based on the detection results, the breath marks were inserted into the text transcripts of both training and test sets, resulting in three TTS models: VITS, VITS w/ baseline, and VITS w/ proposed. These models were trained on two NVIDIA A100 GPUs for 200 epochs, with the same configurations of optimizer and learning rate scheduler as detailed in Section~\ref{breath config}. The batch size was 160 and the peak learning rate was $5 \times 10^{-4}$. 
 During inference, the two key hyper-parameters of VITS, ``noise\_scale\_w'' and ``noise\_scale'' were set to 0.0 and 0.667, respectively. 

{\bf Subjective evaluation settings:}
We conducted two Mean Opinion Score (MOS) tests on Amazon Mechanical Turk. In each test, 40 native English speakers listened to 20 speech samples and rated the naturalness of them on a five-point scale (1 = bad, 2 = poor, 3 = fair, 4 = good, 5 = excellent). The first test (MOS1) assessed the impact of integrating breath detection methods on the overall naturalness of synthetic speech, with an expectation of a positive influence on prosody and rhythm. We randomly sampled 100 synthetic speech utterances from each model for the test, yielding 400 utterances including corresponding ground-truth samples. In this case, not all utterances included breath sounds.
The second test (MOS2) evaluated the effectiveness of breath detection methods in enhancing the naturalness of synthetic breath sounds. We first selected the generated utterances that actually contained breath sounds and then randomly sampled 100 such utterances from each model. During this test, listeners were specifically instructed to pay attention to the breath sounds~\cite{baseline1}.

{\bf Results and analysis:}
The results of the MOS tests are shown in Table~\ref{tab: results of mos}. The following are key findings. 
\begin{itemize}
    


    \item From the MOS1 test results, the scores of VITS and VITS w/ proposed were similar, and VITS w/ baseline showed the lowest MOS value. These results suggested that inaccurate breath detection negatively affected the breath-informed TTS training, whereas our proposed model could overcome the naturalness degradation by using the accurately detected breath marks for the training. Note that, although VITS itself achieved acceptable naturalness, it could not control breath insertion in synthetic speech.
    
    \item From the MOS2 test results, both VITS w/ baseline and VITS w/ proposed achieved higher MOS values than VITS. This indicated that the listeners, who were instructed to pay more attention to breath parts during the test, perceived the synthetic breath by VITS as less natural. We observed that some breath sounds synthesized by VITS tended to be hoarse, but the training with detected breath marks was able to enhance their naturalness. 
    
\end{itemize}

\begin{table}[t]
  \caption{Results of MOS tests. CI: 95\% confidence interval.}
  \label{tab: results of mos}
  \centering
  \vspace{-3mm}
  \begin{tabular}{ l | r r }
    \toprule
    \textbf{Model} & \textbf{MOS1 $\pm$ CI} & \textbf{MOS2 $\pm$ CI}\\
    \midrule
    Ground truth & 4.03 $\pm$ 0.12 & 3.92 $\pm$ 0.13\\
    VITS & 3.35 $\pm$ 0.15 & 3.34 $\pm$ 0.17\\
    VITS w/ baseline & 3.27 $\pm$ 0.15 & 3.50 $\pm$ 0.14\\
    VITS w/ proposed & \textbf{3.37 $\pm$ 0.14} & \textbf{3.55 $\pm$ 0.15}\\
    \bottomrule
  \end{tabular}
\end{table}

\vspace{-5pt}
\section{Discussion}
\vspace{-3pt}
Through multi-speaker training with breath detection and breath mark insertion, we observed that VITS w/ proposed could synthesize breath sounds to some extent for speakers who lacked them in the training data. Specifically, in the ``train-clean-360'' subset, speaker ``3630'' and ``1811'' lacked breath sounds. VITS w/ proposed could synthesize the breath sounds for both speakers, whereas VITS w/ baseline only managed this for ``1811''. We speculate that this was because the baseline model incorrectly detected some silence as breath sounds in the training data of speaker ``3630'', leading the TTS model to misinterpret silence as its breath sounds.

As explained in Section~\ref{tts_experiment}, for the test utterances, breath marks were inserted based on the breath detection results in the ground-truth speech using our trained detection models. However, in actual TTS inference, these breath marks should be predicted by language modeling, similar to tasks like phrasing~\cite{phrasing}. Nevertheless, the significance of our contributions is undiminished, as they achieve accurate breath detection which not only advances current TTS research but also aids researchers in other related fields, such as automatic speaker recognition and speech corpus construction.





\vspace{-5pt}
\section{Conclusion}
\vspace{-3pt}
This study proposes a powerful frame-wise breath detection model, coupled with a detailed analysis of its application in TTS synthesis. By leveraging the annotation from a rule-based approach and employing a self-training method, we have trained the model effectively without manual annotation of training data. The integration of detected breath marks significantly improves the naturalness of synthetic breath sounds.

\bibliographystyle{IEEEtran}
\bibliography{mybib}

\end{document}